\documentstyle[preprint,aps]{revtex}

\input epsf

\begin{document}

\preprint{DUKE-TH-95-97}

\draft

\title{Wave Packet Collisions in Yang-Mills-Higgs Theory}

\author{C. R. Hu, S. G. Matinyan,\thanks{On leave of absence from
Yerevan Physics Institute, Yerevan, Armenia}  B. M\"uller, and D.
Sweet\thanks{Present address:  Physics Department, Boston University,
Boston, MA 02215}}

\address{Department of Physics, Duke University \\
Durham, North Carolina 27708---0305}

\maketitle

\begin{abstract}
We present numerical simulations of colliding wave packets in
spontaneously broken $SU(2)$ Yang-Mills-Higgs theory. Compared with
pure Yang-Mills theory, introducing the Higgs field leads to new
aspects in the dynamics of the system. The evolution of the gauge field
and the Higgs field is investigated as a function of the amplitude of the
wave packets and of the mass ratio of the Higgs and the gauge boson.
We find regions in our parameter space in which initial wave packets
scatter into final configurations with dramatically different momentum
distributions.
\end{abstract}

\pacs{11.15.Kc, 13.85.Hd, 02.60.Cb}

\section{Introduction}

Collisions between classical wave packets have recently been studied
numerically for several interacting relativistic field theories
\cite{TR,GNV,GMMT,OWN}. Interest in this topic arose in connection with
expectations that the rate of multiparticle production processes in
electroweak interactions, which can manifest themselves, e.g., in
baryon number violation, might be unsuppressed at high energies \cite{RE}.

The non-perturbative nature of the baryon number violating amplitude
\cite{HFT} demands a corresponding non-perturbative approach as provided
by semiclassical techniques. The main difficulty in semi-classical
approaches is the treatment of the $2 \to$ many particle transition
amplitude, since the {\it initial} state of high energy particles is
not semiclassical at all, and loop contributions are essential in general
(see, e.g. \cite{VL}). Possible techniques for circumventing this
difficulty of the semiclassical approach have been proposed and studied
in the literature \cite{RST,GHP,RS}.

In the classical approach to scattering, the question is: does there
exist a mechanism for energy transfer from high frequency modes, which
corresponds to two high energy particles, to low frequency modes
representing a multiparticle final state?  At first glance, the answer
to the question, formulated in terms of nonlinear dynamics, seems to be
affirmative since the gauge field theories are nonlinear. However, the
studies of $(1+1)$-dimensional abelian Higgs
model \cite{TR} and  $\lambda\varphi^4$-theory \cite{GNV} have shown
no indication for a non-perturbative
mechanism providing the coupling between the initial
high and the final low frequency modes. For example, in \cite{TR} the
wave packets always passed through each other without being destroyed.
It is important to note that in \cite{TR} the initial states were always
chosen to have small amplitude, which made the nonlinear terms less important.

The important issue here is that the results are strongly influenced
by the nonlinearity due to the non-abelian spin-field coupling, which is
absent in abelian models.  It is this coupling that is responsible for the
infrared instabilities of the pure non-abelian gauge theory.  This can be
seen from the linearized equation describing small perturbations $a_{\mu}^c$
around an $SU(2)$ background field $A_\mu$ (in background gauge):
\begin{equation}
\label{eq1}
(D_{\mu}^2 a_{\nu})^a - 2g\;\varepsilon^{abc}F_{\mu\nu}^ba_{\mu}^c=0~,
\end{equation}
where $D_{\mu}\equiv\partial_{\mu}-ig[A_{\mu}, ~]$ is the gauge covariant
derivative. The second term in (\ref{eq1}) may have any sign. In particular,
this essentially non-abelian coupling drives an instability for
perturbations with isospin polarization orthogonal to the isospin of a
standing wave, which leads to a growth of low frequency modes from initial
high frequency modes  \cite{GMMT}.  This may imply the existence of classical
trajectories of the type required for multiparticle production \cite{GHP},
if this instability persists in more realistic case, e.g., collision of
localized gauge field wave packets.

{}From a more general point of view, the observed inability of the nonlinearity
to furnish a mechanism for the formation of strongly inelastic final states is,
in our opinion, intimately connected with the integrable nature of the
classical
systems considered in \cite{TR} and \cite{GNV}. It is well-known that
non-abelian
 gauge theories are non-integrable in the classical limit , and exhibit
dynamical
chaos \cite{SERGEI,MT} \footnote{Strictly speaking, chaos only
sustains for solutions of finite energy {\it density} \cite{MT}. Finite energy
solutions in $(3+1)$-dimensions will spread out in space at late times and will
therefore
linearize. However, nummerical results indicate that at intermediate times
these
fields generally exhibit exponentially growing perturbations.}. This
dynamical stochasticity of the non-abelian gauge fields together with their
mentioned
dynamical instability are possible sources of the non-perturbative mechanism
for the
coupling between high frequency modes and low frequency modes. At the same
time, it
is important to recall the special role of the Higgs field as a mechanism for
the
suppression of the dynamical chaos of the non-abelian gauge fields
\cite{MATINYAN}.

With this in mind, we studied \cite{OWN} the collision
of two $SU(2)$ gauge field wave packets, homogeneous in the transverse
plane.  As we expected, based on
previous results \cite{GMMT}, the collision of essentially non-abelian
initial configurations trigger the decay of initial high frequency modes
into many low frequency modes with dramatically different momentum
distributions, whereas for abelian configurations (parallel relative isospin
polarizations) wave packets pass through each other without interaction.

The present paper, which studies collisions of wave packets in the $SU(2)$
Higgs model, is a generalization of the earlier work in two directions.
Firstly, it is an extension of \cite{TR} to the non-abelian Higgs model.
Secondly, it is a generalization of the previous work \cite{OWN} to the
case where the $SU(2)$ gauge symmetry is spontaneously broken by a Higgs
field and the fundamental excitations of the gauge field are massive.

One expects that the explicit mass scale introduced by the Higgs field
will act as a cutoff on the low frequency excitations, potentially leading
to drastic changes in the coupling between high and low frequency modes.
We will see that the real situation is more complicated, and the ratio
$\lambda/g^2$ of the Higgs self-coupling $\lambda$ and gauge coupling
$g$ and the vacuum expectation value $v$ of the Higgs field are essential
parameters.

This paper is organized as follows. In Section II, we formulate the
problem. In Section III, we present results from our numerical simulations
and discuss their implications.  Section IV is devoted to an extended
discussion of our results.  In Section V, we conclude and indicate
possible directions for future research.

\section{Formulation Of The Problem}

In this section, we describe scattering of classical wave packets
in the non-abelian Higgs model and its numerical formulation on the
lattice. In particular, we give a brief discussion of the scaling
properties of the classical dynamics. This section is based on previous
work \cite{TR,OWN}.

\subsection{The non-abelian Higgs model}

Here we give a brief discussion on the spontaneously broken $SU(2)$
Yang-Mills theory, in which a charged scalar isodoublet field, the
Higgs field, is coupled to the gauge field. This model retains the
most relevant ingredients of the electroweak theory. The action
describing this model in $(3+1)$-dimensions is given by
\begin{equation}
S = \int {d^3x}{dt}\left[-{{1}\over{2}}{\rm tr}(F_{\mu\nu}F^{\mu\nu})
  + {{1}\over{2}} {\rm tr} \left( (D_{\mu}\Phi)^{\dag} D^{\mu}\Phi
\right) - \lambda \left( {1\over2}{\rm tr}(\Phi^{\dag}\Phi) - v^2 \right)^2
\right]
\end{equation}
with $D_{\mu}=\partial_{\mu}-igA_{\mu}^a \tau^a/2, \;
F_{\mu\nu} \equiv F_{\mu\nu}^a\tau^a/2 = (i/g)\left[D_{\mu}, D_{\nu}
\right]$ and
\begin{equation}
\Phi = \phi^{0}-i\tau^{a}\phi^{a}~,
\end{equation}
where $\tau^{a}$ ($a=1, 2, 3$) are Pauli matrices. Following the
notation of \cite{AAPS}, we represent the complex Higgs doublet by a
quaternion, which is convenient for numerical manipulation. Clearly,
this maintains the correct number of degrees of freedom in the Higgs field.

By a scaling transformation
\begin{equation}
\label{scaling}
x_{\mu}^{\prime}=gv x_{\mu},~~~ \Phi^{\prime}=\Phi/v,~~~
A^{\prime}_{\mu}=A_{\mu}/v~,
\end{equation}
we obtain the action in terms of the primed quantities
\begin{equation}
\label{prime_S}
S = (1/g^2) \int {d^3x^{\prime}}{dt^{\prime}}\left[-{{1}\over {2}}
{\rm tr}(F_{\mu\nu}^{\prime}F^{\prime\mu\nu}) + {1\over2}{\rm tr}
\left( (D^{\prime}_{\mu}\Phi^{\prime})^{\dag}D^{\prime\mu}\Phi^{\prime}
\right)- \lambda^{\prime} \left({1\over2}{\rm tr}(\Phi^{\prime\dag}
\Phi^{\prime}) - 1\right)^2\right]
\end{equation}
with $\lambda^{\prime}=\lambda/g^2$.

Within classical physics, the prefactor ${{1}/{g^2}}$ in (\ref{prime_S})
is irrelevant, leaving $\lambda^{\prime}=\lambda/g^2$ as the only relevant
parameter in the action. Note that $\lambda^{\prime}$ is proportional to
the square of ${M_H}/{M_W}$, the mass ratio of the Higgs and $W$-boson.

The elementary excitation modes $\rho$ and $W_{\mu}$ are best described
in the unitary gauge
\begin{eqnarray}
\Phi &=& (v+\rho/\sqrt{2})U(\theta)~, \\
A_{\mu} &=& U(\theta)W_{\mu}U^{-1}(\theta) -
(1/ig)\left(\partial_{\mu}U(\theta)\right) U^{-1}(\theta)~,
\end{eqnarray}
where $U(\theta)=\exp(i\tau^{a}\theta^{a})$, $\rho$ describes oscillation
of the Higgs field about its
vacuum expectation value, and $W_{\mu}$ is the field of the gauge boson.
$W_{\mu}$ and $\rho$  obey the following classical equations of motion
\begin{eqnarray}
[D_{\mu}, F^{\mu\nu}]+M_W^2 W^{\nu}+{{1}\over{\sqrt{2}}}g^2 v\rho W^{\nu}
+{{1}\over{4}}g^2\rho^2 W^{\nu} &=& 0~,      \label{w_equation} \\
(\partial_{\mu}\partial^{\mu}+M_H^2)\rho+3\sqrt{2}\lambda v \rho^2
+\lambda\rho^3-{{\sqrt{2}}\over{4}}g^2 v W^{a}_{\mu}W^{a \mu} -
{1\over 4}g^2\rho W^{a}_{\mu}W^{a \mu} &=& 0~,             \label{h_equation}
\end{eqnarray}
where $M_H=2v\sqrt{\lambda}$ and $M_W=gv/\sqrt{2}$. $D_{\mu}$ and
$F^{\mu\nu}$ are defined in terms of $W_\mu$. After a scaling
transformation similar to that in (\ref{scaling}), it is easy to see
that the above equations of motion depend on a single parameter: the
mass ratio $M_H/M_W$. However, in the simulation of colliding wave packets,
there are other parameters involved in the initial condition.

\subsection{Scattering of classical wave packets}

Our numerical study is based on the Hamiltonian formulation of lattice
$SU(2)$ gauge theory \cite{CRUK} (see \cite{OWN,GT} for more
details), in which the dynamic variables are link variables defined as
\begin{equation}
U_{\ell} ~=~ \exp(-igaA^c_{\ell}\tau^c/2)~,
\end{equation}
where $\ell$ stands for the link index.  As in \cite{OWN}, we work on a
one dimensional lattice with a physical size $L=Na$, where $N$ is the
number of lattice sizes and $a$ the lattice spacing. We arrange initially
two Gaussian wave packets with average momenta ${\bf k}=(0,0,{\bar k})$,
and width $\Delta k$. Our goal is to simulate the collision of two $W$-boson
wave packets in the background of the Higgs condensate.

Before actually constructing the wave packets, one has to deal with the
gauge fixing problem. In the Hamiltonian formulation of lattice gauge
theory, temporal gauge $A_0=0$ is most convenient.  On the other hand, one
must construct the $W$-boson wave packets in the unitary gauge and then
transform back to obtain the initial conditions in the temporal gauge.
The gauge field for a configuration of two well-separated wave packets
in the unitary gauge can be written as
\begin{equation}
W^{c,\mu}=W_{L}^{c,\mu}+W_{R}^{c,\mu}~,
\end{equation}
with $c$ being the isospin index and $\mu$ the Lorentz index.
$W_{L}^{c,\mu}$ is a left-moving wave packet, initially centered at
$z_L$;  $W_{R}^{c,\mu}$ is a right-moving wave packet, initially
centered at $z_R$. In our simulation, $z_L$ and $z_R$ are chosen in a
way such that the two wave packets are positioned symmetrically about
the center of the lattice. Specifically, we take transversely polarized
wave packets
\begin{eqnarray}
W_{L}^{c,\mu} &=& (0, 0, 1, 0) n^{c}_{L}\psi (z-z_L, -t)~,  \\
W_{R}^{c,\mu} &=& (0, 0, 1, 0) n^{c}_{R}\psi (z-z_R, +t)~,
\end{eqnarray}
with $n^{c}_L$ and $n^{c}_R$ being the polarization vectors in isospin
space. We choose $n^{c}_R=(0,0,1)$ fixed and leave $n^{c}_L$ free to be
varied. The above choice satisfies the relation $\partial_{\mu}W^{\mu}=0$.

Because we have chosen transversely polarized wave packets\footnote{Note
that under realistic conditions the luminosity for transversely polarized
gauge bosons in proton-proton system is typically two orders of magnitude
higher than for longitudinally polarized ones and increases with energy
\cite{Dawson}.}, they already satisfy the temporal gauge condition $W_0=0$.
If we had chosen, instead, longitudinally polarized wave packets (as in ref.
\cite{TR}), we would have needed to apply a gauge transformation
$U({\bf x},t)$ which transforms $W_{\mu}$ to $A_{\mu}$
\begin{equation}
A_{\mu}=U W_{\mu} U^{-1} - (1/ig) (\partial_{\mu}U)U^{-1}~,
\end{equation}
such that $A_0=0$, i.e.,
\begin{equation}
\partial_0 U = igU W_0~.
\end{equation}
In our case, obviously, $U=1$.

To construct the wave packets, we need to specify the functional form of
$\psi ({\bf x}, t)$. A right-moving wave packet centered at $z=0$ at
time $t=0$ with mean wave number $\bar k$, width $\Delta k$, and mean
frequency ${\bar \omega}=\sqrt{{\bar k}^2 + M_W^2}$ is described by
\begin{equation}
\psi (z, t) = {{\sqrt{\hbar}}\over{\sqrt{4\pi\sqrt{\pi}\Omega
\Delta k \sigma}}}\int_{-\infty}^{\infty} dk_z
e^{-(k_z-{\bar k})^2/{2(\Delta k)^2}} \left[ e^{i(\omega (k_z) t-k_z z)}
+ c.c. \right]~,
\end{equation}
with $\omega (k_z)=\sqrt{{k_z}^2+M_W^2}$  and
\begin{equation}
\Omega= {\bar \omega} \left[ 1+{1\over 4} \Big[ 1- \left( {\bar k
\over \bar \omega}\right)^2\Big] \left({\Delta k\over\bar \omega}\right)^2
+ {\cal O} \left( \Big( {\Delta k\over\bar \omega} \Big)^4 \right)\right]~,
\end{equation}
where the amplitude of the wave packet is fixed by requiring energy
equal to $\hbar {\bar \omega}$ (``one particle'') per cross sectional area
$\sigma$. In the following we will set $\hbar = 1$.

Performing the $k_z$-integral at $t=0$ gives
\begin{equation}
\label{init_phi}
\psi |_{t=0} = \sqrt{{2\Delta k\over\sqrt{\pi}\Omega\sigma}}
e^{-(\Delta k z)^2/2}\cos({\bar k}z)~.
\end{equation}
Since the differential equations are of second order in time, one also
needs to specify $\dot{\psi}\equiv \partial \psi/\partial t$ at $t=0$, which
is found to be
\begin{eqnarray}
\dot{\psi} |_{t=0}&=& {i\over\sqrt{4\pi\sqrt{\pi}\Omega
\Delta k \sigma}} \int_{-\infty}^{\infty} dk_z \omega (k_z)
e^{-(k_z-\bar k)^2/2(\Delta k)^2} \left[ e^{i(\omega (k_z) t-k_z z)}
-c.c.\right]_{t=0} \nonumber \\
&=& \sqrt{2\bar \omega\Delta k\over\sqrt{\pi}\sigma (\Omega/\bar\omega)}
e^{-(\Delta k z)^2/2} \left[\sin(\bar kz) +{\bar k\over\bar\omega}
{\Delta k\over\bar\omega}\Delta k z \cos(\bar kz) \right.
\nonumber \\
&&\qquad  \left.  +{1\over2} \left[ 1- \left({\bar k\over\bar\omega}\right)^2
\right] \left({\Delta k\over\bar\omega} \right)^2 [1-(\Delta k z)^2]
\sin(\bar kz)+{\cal O}\left( \Big({\Delta k\over\bar \omega}\Big)^3 \right)
\right]~.
\end{eqnarray}

Furthermore, the initial condition for the Higgs field is chosen as the
vacuum solution:
\begin{equation}
\phi ^0=v, ~~~ \phi ^a=0, ~~~ \dot{\phi ^0}=\dot{\phi ^a}=0  ~~~~~
{\rm at} ~~ t=0~.
\end{equation}

To determine the number of independent parameters, we make use of the
scaling transformation (\ref{scaling}) for $\psi |_{t=0}$. In terms of the
primed quantities, equation (\ref{init_phi}) reads
\begin{eqnarray}
\psi^\prime|_{t^\prime =0} &=& (1/v)\psi|_{t=0} \nonumber \\
&=& (1/\pi^{1/4})\sqrt{{\Delta k/M_W\over(\Omega/M_W)
(\sigma M_W^2/g^2)}} e^{-(\Delta k/M_W)^2 z'\,^2/4}
\cos[(\bar k/M_W)z^\prime/\sqrt{2}] ,  \label{amplitude}
\end{eqnarray}
where
\begin{equation}
{\Omega\over M_W}=\sqrt{1+(\bar k/M_W)^2} \;
\left[ 1+{1\over4}{(\Delta k/M_W)^2 \over[1+(\bar k/M_W)^2]^2} +
{\cal O} \Big( (\Delta k/\bar k)^4 \Big) \right]~.
\end{equation}
The above initial condition contains three dimensionless parameters:
${{\bar k}}/{M_W}$, ${\Delta k}/{M_W}$, and ${\sigma M_W^2}/{g^2}$.
There appears one more parameter in the initial condition, i.e., the
relative rotation in isospin space between the two wave packets, which
we denote by $\theta_c$.

Combining equations of motion and initial condition, our ansatz has
five independent parameters: ${M_H}/{M_W}$, ${\bar k}/{M_W}$,
${\Delta k}/{M_W}$, ${{\sigma M_W^2}/{g^2}}$, and $\theta_c$. The
parameter ${\bar k}/{M_W}$, which sets the energy of collisions in units
of the $W$-boson mass, is referred to as the energy parameter.
${\Delta k}/{M_W}$, whose inverse specifies the width of each wave packet
in position space, can be called the width parameter. The amplitude of
each wave packet depends on ${\bar k}/{M_W}$ as well as
${\Delta k}/{M_W}$, but more crucially, on ${\sigma M_W^2}/{g^2}$. In
our simulation, we always require $\bar k \gg \Delta k$ so that the
wave packets are well-defined objects. Furthermore, we choose
$\bar k \gg M_W$ to model high energy scattering.

In our numerical calculations, the $SU(2)$ coupling constant $g$ was fixed
to be $0.65$. However, due to
the scaling properties of the equations of motion and the initial
conditions, the results of our calculation do not depend on the
particular choice of $g$ and $v$.  This can be verified by the fact
that the amplitude of the wave packets only depends on the ratio of
$\sigma$ and $g^2$.  Also, since the dynamics does not depend on a
particular choice of $\sigma$ or $g$ as long as the ratio
$\sigma/g^2$ is fixed (assuming that $M_W$ and other parameters remain
fixed), we can predict, from the result for one coupling $g$ at a
certain value of $\sigma$, the result for another coupling $g'$ at a
different $\sigma' = (g'/g)^2\sigma$.  Hence, a change of the value for
the gauge coupling $g$ simply corresponds to a rescaling of the
parameter $\sigma$ controlling the amplitude of initial wave packets.

\section{Numerical Results}

\subsection{Dependence on the mass ratio $M_H/M_W$}

As established in the previous work \cite{OWN}, the behavior of the
wave packet collisions is governed by the nonlinearity due to the
self-interaction of the gauge field. For two wave packets of parallel
isospin polarizations in the pure Yang-Mills theory, where the nonlinear
self-coupling in the gauge field is absent, we found no indication of
any interaction \cite{OWN}. This provided a check on our numerical
procedure and showed that the artificial interactions introduced by the
formulation in terms of compact lattice gauge fields did not affect the
results.

Here, for the Yang-Mills-Higgs system, the situation is more involved.
Besides the nonlinearities due to the gauge field self-interaction, there
exist other nonlinearities induced by the gauge-Higgs coupling and by the
Higgs self-interaction.  However, in the case where most energy remains
contained in the gauge field and the Higgs field is only slightly excited,
one can expect that the gauge field self-interaction will be the major
contributor to the nonlinearities observed in the system.  Note that the
amplitudes of the gauge and Higgs fields shown in the figures below are
in the temporal gauge.  This means that the longitudinal part of the gauge
field is not fully represented in the figures.

The top rows of Figure 1 and Figure 2 show a few ``snapshots'' of the
space-time development of the colliding $W$-boson wave packets with
$M_H=M_W=0.126$, $\sigma=0.336$, $\bar k =\pi/5$, and $\Delta k=\pi/100$ for
parallel (Figure 1) and orthogonal (Figure 2) isospin polarization,
respectively. The figures show the absolute magnitude of the scaled gauge field
amplitude, $|{\bf A}^\prime|=|{\bf A}|/v$.
For parallel isospin orientations, the
result of the ``collision'' is a slight distortion of the initial wave
packets showing no sign of significant inelasticity. In contrast, the
top row of Figure 2 illustrates that the collision of two wave packets
with orthogonal relative polarizations in isospin space is strongly inelastic.

The difference between the two figures is even more clearly illustrated
by looking at the evolution of the absolute value of the Fourier
transform of the gauge invariant energy density (scaled by $v^2$)
\begin{equation}
\label{spectrum}
\widetilde{{\cal E}}\left({\bf k}, t\right)=\frac{{\cal E}\left({\bf k},
t\right)}{v^2} ~=~
\frac{1}{4 v^2}\,\left|\,\int {d^3x}~e^{i{\bf k}\cdot{\bf x}}~{\rm Tr}~
\left[\, {\bf E}^2\left({\bf x}, t\right) ~+~
{\bf B}^2\left({\bf x}, t\right) \,\right]\,\right|~,
\end{equation}
where ${\bf E}$ is the gauge electric field and ${\bf B}$ the gauge magnetic
field. It is seen that the spectrum for the parallel isospins (median
row in Figure 1) does not change its shape dramatically, while for the
case of orthogonal isospins (median row in Figure 2), the spectrum spreads
out widely. The spike at $k=0$ in these spectra corresponds to the total
energy contained in the transverse gauge field. From its slight decrease
in Figure 2, we see that the energy transferred to the Higgs and the
longitudinal gauge fields during the collision is small ($\approx 10\%$).
This reflects the fact that the nonlinearity due to the gauge field
self-coupling dominates. Furthermore, the bottom rows of Figures 1 and 2
illustrate the time evolution of the Higgs field excitations\footnote{As
seen from (\ref{h_equation}) oscillations of the gauge boson field act as
a source for Higgs field excitations.  The equation (\ref{w_equation}) for
the gauge field does not possess a source term.} around its condensate
value $v$ (which is scaled to unity) accompanying the collision process
shown in the top rows of Figures 1 and 2.  Here we have plotted the square
$|\Phi^\prime|^2=|\Phi|^2/{v^2}$ of the Higgs field as a function of space
coordinate at three different times. Throughout our simulations, we have kept
$k_{min}\ll\Delta k\ll\bar{k}\ll k_{max}$, where $k_{min}$ and $k_{max}$ are
the minimum and maximum momentum on the lattice, respectively. This
ensures that the wave packets
are smooth on the lattice. But during collisions, unlike
quantum mechanics, classical dynamics does not provide a mechanism for
stopping power from flowing to
very low frequency modes (close to $k_{min}$) or to very high frequency modes
(close to $k_{max}$). In our calculations, the power flowing to high frequency
modes does not cause a deterioration in the local smoothness of the gauge field
at the end of the simulations.

To display dependence on the mass ratio $r={{M_H}/{M_W}}$, the
top row of Figure 3 shows the collision of two orthogonally polarized
$W$ wave packets at the final time $(t=580)$ for three different values of
$r={M_H/ M_W}$. It is seen that the ``inelasticity'' is more pronounced
for small $r$. On the other hand, the distortions in the wave packets
still survive at large $r$ (even at $r=100$, not shown here). This can
be understood as follows.  Remember that there are three sources of
nonlinearity, namely, gauge field self-coupling, gauge-Higgs coupling,
and Higgs self-coupling.  As the Higgs mass increases, the Higgs modes
begin to decouple. As a result, the interaction between gauge and Higgs
field diminishes and hence contributes less to the nonlinear effects.
The gauge field self-interaction is not affected by the change in the
Higgs mass and acts as the main contributor of nonlinear effects observed
during the collision. The median row of Figure 3 is the analogue of
the top row of Figure 3 in momentum space, as defined in (\ref{spectrum}).
Again, it gives a clearer picture of the inelasticity of the collision
process.  The bottom row of Figure 3 demonstrates that the amplitude of
the Higgs field excitations becomes smaller as the Higgs mass is increased,
while their frequency increases with $M_H$.

It is remarkable that for small Higgs mass, as seen for $r=0.1$ in Figure 3,
the Higgs field oscillates not about the vacuum expectation value $v$ but
rather
about zero. The
observed behavior holds even at larger values of $r$, up to $r \approx 0.5$
(not
shown here). This suggests that for not too large $r$, the collision of gauge
field wave packets, accompanied by energy transfer from gauge field to Higgs
field,
leads to restoration of the broken symmetry. This phenomenon occurs for gauge
field configurations with large amplitude \footnote{The idea that the
restoration of
vacuum symmetry is possible in the background of intense gauge fields was first
noted
in \cite{KLINDE}. Also see \cite{KPC}), where the role of external gauge fields
in the
restoration of broken symmetries was considered.}. Indeed, it is easy to see
that
$\rho=-\sqrt{2}v$ (i.e. $|\Phi|=0$) is an exact solution to equation
(\ref{h_equation}).
Inserting $\rho=-\sqrt{2}v$ into equation (\ref{w_equation}) leads to the pure
Yang-Mills
equation for massless
$W$-bosons. In terms of excitations around this state, $\chi=\rho+\sqrt{2}v$,
we rewrite
equations (\ref{w_equation}) and (\ref{h_equation}) as
\begin{eqnarray}
[D_{\mu}, F^{\mu\nu}]+\frac{1}{4}g^2\chi^2 W^{\nu} &=& 0~,
\label{wprime_equation} \\
\left[\partial_{\mu}\partial^{\mu}-M_H^2(1+\frac{g^2W^2}{8\lambda v^2})\right]
\chi
 + \lambda\chi^3 &=& 0~,      \label{chi_equation}
\end{eqnarray}
where $W^2=-(W_i^a)^2 < 0$ for transverse polarized wave packets (the sum over
spatial index
$i$ and isospin index $a$ is assumed here and in below). After dropping a
constant term
$\lambda v^4$, the corresponding effective potential describing the excitations
$\chi$ has the following form:
\begin{equation}
V(\chi,~W_{\mu})=-\lambda v^2(1-\eta)\chi^2+\frac{\lambda}{4}\chi^4~,
\label{potential}
\end{equation}
where we denote
\begin{equation}
\eta \equiv \frac{g^2(W_i^a)^2}{8\lambda v^2}
 = \frac{1}{r^2} \left( \frac{W_i^a}{v} \right)^2~.     \label{ETA}
\end{equation}
In the following we use $\eta$ as a parameter in which the true
intensity $(W_i^a)^2$ of the high frequency gauge field pulses is replaced by
its
space-time average $\langle W^2 \rangle$. Depending on whether $\eta<1$ or
$\eta > 1$,
the potential (\ref{potential}) has two different {\it stable} minima:
\begin{eqnarray}
{\rm for}~ \eta<1, &~& \chi_{min}=\pm\sqrt{2}v(1-\eta)^{1/2}~, ~~~{\rm i.e.}~~~
|\Phi|=v(1-\eta)^{1/2}~,  \\
{\rm for}~ \eta \geq 1, &~&\chi_{min}=0~, ~~~{\rm i.e.}~~~ |\Phi|=0~.
\end{eqnarray}
Stable excitations about these ``vacua'' have the following squared masses:
\begin{eqnarray}
{\widetilde{M}}_W^2 &=& M_W^2(1-\eta) \theta(1-\eta)   \label{WMASS}  \\
{\widetilde{M}}_H^2 &=& \frac{M_H^2}{2}(1-\eta) [1+\theta(1-\eta)]
\label{HMASS}
\end{eqnarray}
Thus for $\eta > 1$ , the broken symmetry is restored and oscillations of the
scalar
field occur about the symmetrical state $|\Phi|=0$, not about $|\Phi|=v$. The
effective
mass of the gauge bosons in the region where the symmetry is restored vanishes.
For $\eta<1$,
the ratio between the effective masses ${\widetilde{M}}_H$ and
${\widetilde{M}}_W$ has
no dependence on $\eta$ and remains $r=M_H/M_W$. Relations
(\ref{WMASS}) and (\ref{HMASS}) are characteristics of a second order phase
transition.
The expression (\ref{ETA}) for $\eta$ shows that in the regime of large
$(W_i^a)^2 > v^2$,
this phase transition can occur for experimentally favorable mass ratio $r>1$.

Oscillations of the scalar field around the new symmetrical minimum $|\Phi|=0$
are clearly
seen for $r=0.1$ where $\eta \approx 22.3 $ (see Figure 3). These numerical
results provide indications for transition from the phase with spontaneously
broken
$SU(2)$ symmetry and asymmetric vacuum to the phase with restored $SU(2)$
symmetry
and symmetric vacuum as a result of the collisions.
Since $\eta$ is the only relevant parameter in question here,
this transition can occur either for small $\lambda$ (light Higgs) or for large
$\lambda$
(heavy Higgs) if amplitude of the gauge field is large enough. In Figure 3, it
is also
interesting to notice that the spatial region showing symmetry restoration
seems to be
wider than the region
where the colliding wave packets stay visibly large.

It is clear that the phenomenon discussed above does not depend on the
one-dimensionality
of space in our calculations. However, it is to be expected that the symmetry
restoration
would not persist as long in three dimensions as the wave packets disperse more
rapidly after
the collision causing the squared amplitude $\langle W^2 \rangle$ to decay more
rapidly.

\subsection{Yang-Mills and BPS limit}

In the light of the previous work \cite{OWN}, it is instructive to study
the limiting case of the present system as $M_H, ~ M_W
\rightarrow 0$ while fixing $r={{M_H}/{M_W}}$ (which is chosen
to be $1$ here). Of course, this corresponds to the limit $v \rightarrow 0$,
where one expects that the gauge field in the Yang-Mills-Higgs system behaves
most closely to that in a pure Yang-Mills system. In Figure 4, the
top row shows snapshots of the collision of two orthogonally polarized
wave packets with $M_H=M_W=0.001$ and the bottom row exhibits the
corresponding spectra.  Qualitatively, these figures show that the time
evolution is very similar to that seen in the pure Yang-Mills system
(see Figures 2 and 4 in \cite{OWN}).

The second interesting limit is the Bogomolny-Prasad-Sommerfield (BPS)
limit \cite{BPS} where $M_H\rightarrow 0$  but $M_W$ is finite
($\lambda\rightarrow 0$, $v$ fixed). Figure 5 shows the snapshots and
spectra for this case with $M_W=0.126$ and $r=0.01$.  Again, we display
only the orthogonal case, which reveals complete destruction of the wave
packets as in the pure Yang-Mills limit. It is interesting to note that
in this limit the static force between equally charged $W$-bosons
vanishes due to the precise cancellation between the photon and (massless)
Higgs exchange diagrams, which is a result of the electromagnetic duality
\cite{MO}. For a pair of oppositely charged $W$-bosons, the contributions
from these diagrams add to each other, doubling the attraction.

\subsection{Dependence on the initial amplitude and energy}

In our simulation, the most crucial role is played by the initial
amplitude of the wave packets. As in the pure Yang-Mills case \cite{OWN},
we find that the amount of ``inelasticity'' observed in the present
system is closely linked to the magnitude of the dimensionless amplitude
(\ref{amplitude}).  This amplitude  depends on several independent
parameters: $\sigma M_W^2 / g^2$, ${\Delta k}/ M_W$, and
${\Omega}/{M_W}\approx {\bar k}/{M_W}$ (for ${\bar k}\gg M_W$),
each of which has a different physical meaning.  Noting that the gauge
coupling constant is fixed to be $g=0.65$ throughout our simulation
and $M_W$ is a fixed quantity in reality, the best way to study the
amplitude dependence is to vary $\sigma$ without changing anything else.
By varying $\sigma$, we find that the nonlinear effects increase with
amplitude.  For a very large $\sigma$ (very small amplitude), we find
no indication of ``inelasticity'' in the colliding wave packets at the
end of the simulation.

To search for the energy dependence, we have to study the dependence on
$\bar k$, which determines the energy of the initial wave packet. In
the meantime, we fix $\sigma$, $M_W$, and $M_H$.  $M_W$
and $M_H$ are chosen to be much smaller than $\bar k$ to model high
energy scattering.  Furthermore, as we change $\bar k$, $\Delta k$ is
either fixed or changed proportionally to fix the ratio ${\Delta k}/{\bar k}$.
In Figure 6, we display snapshots of collisions at the final time for
different sets of $\bar k$ and $\Delta k$.  Figure 7 shows the
corresponding spectra. The orthogonal isospin cases in Figures 6 and 7
are shown in the left column with their parallel isospin counterparts in the
right column.

Clearly, the observed nonlinear effects in the orthogonal cases are
qualitatively similar for different $\bar k$ or $\Delta k$; while in
the parallel cases, regardless of $\Delta k$, the nonlinear effects
disappear as $\bar k$ is increased from $\pi/25$ to $\pi/5$.

\section{Discussions}

\subsection{Amplitude Dependence}

Our numerical calculations show for a wide range of parameters that
the wave packet collisions with orthogonal isospin orientation are
strongly inelastic if the scaled initial amplitude (see equation
(\ref{amplitude}), $k\gg M_W$)
\begin{equation}
\left({2\Delta k\over\sqrt{\pi}\bar k\sigma v^2}\right)^{1/2} \label{24}
\end{equation}
is of the order of unity.  We recall that the expression (\ref{amplitude})
for the scaled amplitude was determined by the condition that the wave packet
contains one particle per transverse area $\sigma$.

Although, in the strict sense, our configurations describe wave packets
which are infinitely extended in the transverse direction and hence
contain infinitely many particles, only a finite transverse area
influences the dynamics over a finite period of time.  As argued in
\cite{OWN}, causality restricts that area to $\sigma (T_s) = \pi T_s^2$
where $T_s$ is the elapsed time after the impact of the two
wave packets.  The relevant number of particles in the initial state is
therefore given by
\begin{equation}
N_i^{\rm eff} = {\sigma(T_s)\over\sigma} = {\pi T_s^2\over \sigma}.
\end{equation}

What is the lower bound on $N_i^{\rm eff}$ under realistic conditions,
for which strongly inelastic events occur?  Let us first estimate the
constraints on our parameters from a realistic point of view.  Clearly,
we must have $\bar k\gg v$.  The natural spread of any $W$-boson wave packet
produced in high energy interactions is of order $\Delta k\sim v$ in
the comoving reference frame.  Therefore the typical transverse area
of the $W$-boson wave packet is of order $\sigma \approx 1/v^2$.  Due
to Lorentz contraction, its longitudinal momentum spread will generally
be much larger than $v$, or of order $\Delta k_{\Vert}\sim \gamma
v$, where $\gamma \approx \bar k/M_W$ is the Lorentz factor.  As a result,
$\Delta k_{\Vert}/\bar k$ will be approximately independent of the
collision energy, with a value not much smaller than one.  Assuming,
e.g., $\Delta k/\bar k \approx 0.5$ in (\ref{24}), we obtain an amplitude of
order unity implying that a few $W$-bosons per area $\sigma$ in the
initial state could produce strong inelasticity.  Of course, the
precise lower bound on the particle number will depend on the detailed
shape of the wave packets and requires a full three-dimensional
analysis.  But our estimate shows that strongly inelastic events are
not excluded for collisions of wave packets containing few particles.
In this respect, the results of our analysis correspond to those of
Singleton and Rebbi \cite{RS} who found that few-particle initial states
may not be excluded for baryon number nonconserving processes
resulting in multi-particle final states.

As mentioned above, the finite transverse size of order $v^{-1}$
limits the applicability of our calculation for the real
three-dimensional case to times $T_s \le v^{-1}$.  Since the
inelasticity is clearly revealed for times $T_s\sim 100$ (see e.g.
Figure 5), our numerical results apply most confidently to very high
energy where $\bar k/v \ge 10^2$.

\subsection{Energy Dependence}

We now turn to the question of the energy dependence of the nonlinear
effects seen in the wave packet collisions.  In the $(1+1)$-dimensional
abelian Higgs model, the nonlinearities were clearly found to decrease
with energy \cite{TR}. For the non-abelian Higgs model discussed here,
the answer is given in Figures 6 and 7.  The inelasticity seen in the
orthogonal isospin cases does not change significantly with energy,
while it dies out in the parallel isospin cases as $\bar k$ increases.
This shows the fundamental role of the non-abelian nature of the
$W$--$W$ interaction in the formation of strongly inelastic final states.

{}From Figures 6 and 7 one can also see that the inclusion of the Higgs field
produces new phenomena which are not seen in the pure Yang-Mills
system: For initial configurations with parallel isospin, in which case
nonlinear interactions of the gauge bosons are absent, lowering of
the parameter $\bar k$ leads to inelastic final states. This is
exclusively due to the Higgs field since the collision of wave packets
with parallel isospin configurations in the pure Yang-Mills theory
always leads to  elastic final states independently of $\bar k$ \cite{OWN}.
Of course, this pattern indicates that the influence of the Higgs coupling
to the gauge field increases for smaller $\bar k$. How does one understand
this behavior? For this purpose, we recall that all our calculations
are in the regime of high energy ($\bar k \gg v, M_W$).  For the highest
energy of the collisions ($\bar k =\pi/5$) where parallel polarized
wave packets scatter elastically, one may think that the transversely
polarized $W$--$W$ scattering (elastic in these collisions) proceeds via
exchanges of the gauge and of the Higgs bosons in the tree approximation.
The first contribution prevails at high energy, but it does not contribute
to the inelastic final states for the parallel isospin orientation.
Inelasticity may arise here only from the nonlinear coupling of the gauge
and Higgs fields or, in diagrammatic language, due to the Higgs exchange
whose contribution increases with the lowering of $\bar k$.

\section{Conclusion}

We have numerically studied collisions between classical wave packets
of transversely polarized gauge bosons in the spontaneously broken
Yang-Mills-Higgs theory.  Our main results are the following:

\begin{enumerate}
\item We have found evidence for the creation of final states with
dramatically different momentum distributions (strongly ``inelastic''
events) for a wide range of the essential parameters.

\item These inelastic events persist at the highest investigated energies
$(\bar k/M_W \sim 10^2)$ for collisions with orthogonal isospin
polarization, reflecting the essentially non-abelian character of the
interaction.  For parallel isospin configurations, in contrast, the
inelastic events, which are solely due to the Higgs field, occur only for
lower energies.

\item Under more realistic conditions as discussed in Section IV.A, the
inelastic
events are not excluded for initial configurations with few particles.

\item We have observed, at least for $r \leq 0.5$ (with fixed amplitude for
the gauge field wave packets),  the phenomenon of symmetry restoration as a
result of the wave packet collisions. This transition from the asymmetrical
state to
the symmetrical one is
governed by a single parameter $\eta$, which depends on both the mass ratio $r$
and
the amplitude of the wave packets.
\end{enumerate}

Summarizing, we conclude that the introduction of the Higgs field (in
the broken symmetry phase) does not in general spoil the inelasticity
of the final state in collisions with orthogonal isospin orientation.
Our results provide a strong motivation for exploring related
phenomena in $(3+1)$ dimensions.  This would allow one to study
collision between realistically shaped wave packets and investigate the
particle number content of inelastic final states.  Last, but not
least, it would be interesting to study the winding number change
associated with these collisions.


\acknowledgements

We thank S. D. H. Hsu for discussions and useful comments. Especially, we
want to thank the referee for numerous constructive comments and suggestions,
which were very useful in preparation of a revised version of this paper.
This work was supported, in part, by grants DE-FG05-90ER40592 and
DE-FG02-96ER40945 from the U.S. Department of Energy, and by the North Carolina
Supercomputing Center.

\begin{figure}
\caption{Collision of two $W$ wave packets with parallel isospin
polarizations.  We choose $M_H=M_W=0.126$, $\bar k = \pi/5$,
$\Delta k = \pi/100$, $g=0.65$, and $\sigma = 0.336$. This simulation,
as well as all others below, was performed on a lattice of
length $L = 2048$ and lattice spacing $a = 1$.  The top row shows the
space-time evolution of the scaled gauge field amplitude $|{\bf A}^\prime|$,
the median row exhibits the corresponding Fourier spectra
of the gauge field energy density, eq. (\protect{\ref{spectrum}}), and the
bottom row
shows the space-time evolution of the scaled Higgs field
$\vert{\Phi^\prime}\vert^2$.
The abscissae of top and bottom rows are labelled in units of the
lattice spacing, and the abscissa of the median row is in units of
$\pi /1024$.}
\label{Figure1}
\end{figure}

\begin{figure}
\caption{Same as Figure 1, but for orthogonal isospin polarizations.}
\label{Figure2}
\end{figure}

\begin{figure}
\caption{Mass ratio $r=M_H/M_W$ dependence of the collisions shown for
three different values of $r$ at the end of our calculation $(t=580)$.
Here we have used orthogonally polarized $W$ wave packets.  Except the
mass of the Higgs, all the other parameters are fixed to be the same as
in Figure 1.}
\label{Figure3}
\end{figure}

\begin{figure}
\caption{Collision of two orthogonally polarized wave packets in the
Yang-Mills limit.  Except for $M_H=M_W =0.001$, all the other
parameters are the same as in Figure 1.  The top row shows the space-time
evolution of the scaled gauge field amplitude. The bottom row shows the Fourier
spectra of gauge field energy density, as defined in (\protect\ref{spectrum}).}
\label{Figure4}
\end{figure}

\begin{figure}
\caption{Collision of two orthogonally polarized wave packets in the
BPS limit.  Here we fix $M_W=0.126$ but choose a small mass for the
Higgs: $M_H=10^{-2} M_W$.  All the other parameters are the same as in
Figure 1.  The top row shows the space-time evolution of the scaled gauge field
amplitude. The bottom row shows the Fourier spectra of gauge field energy
density,
as defined in (\protect\ref{spectrum}).}
\label{Figure5}
\end{figure}

\begin{figure}
\caption{Final states $(t=1100)$ of the scaled gauge field for three different
sets of $\bar k$ and $\Delta k$ are shown.  Here $\sigma=0.504$, $M_H=M_W =
0.01$, and all the other parameters are the same as in Figure 1.  Left column:
orthogonal isospin
orientations. Right column: parallel isospin orientations.}
\label{Figure6}
\end{figure}

\begin{figure}
\caption{Energy spectra for the cases shown in Figure 6.}
\label{Figure7}
\end{figure}

\newpage

\ifx\pansw\pictures
\def\epsfsize#1#2{0.95#1}
\centerline{\epsfbox{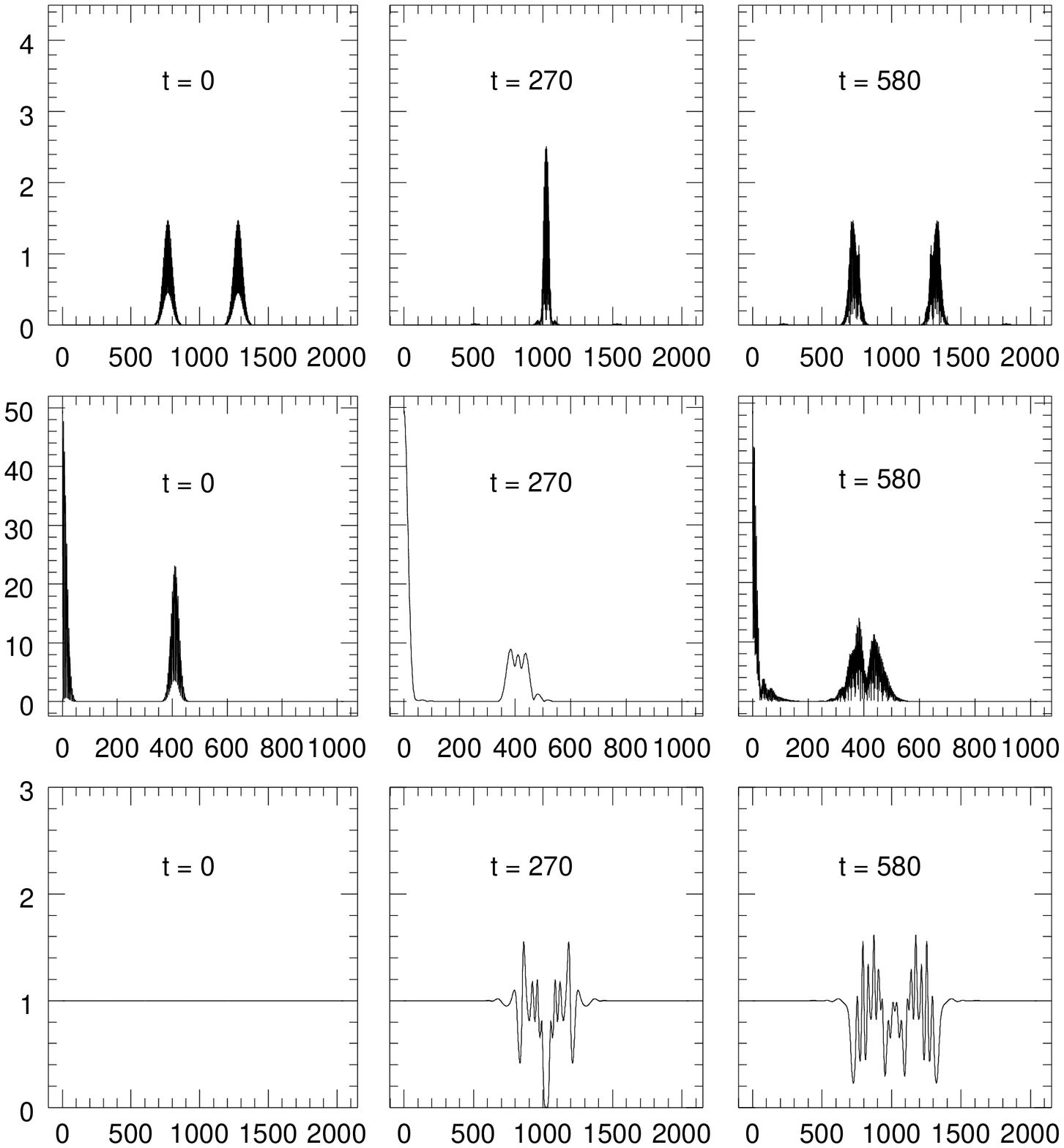}}
\centerline{Fig.1}
\else\fi

\ifx\pansw\pictures
\def\epsfsize#1#2{0.95#1}
\centerline{\epsfbox{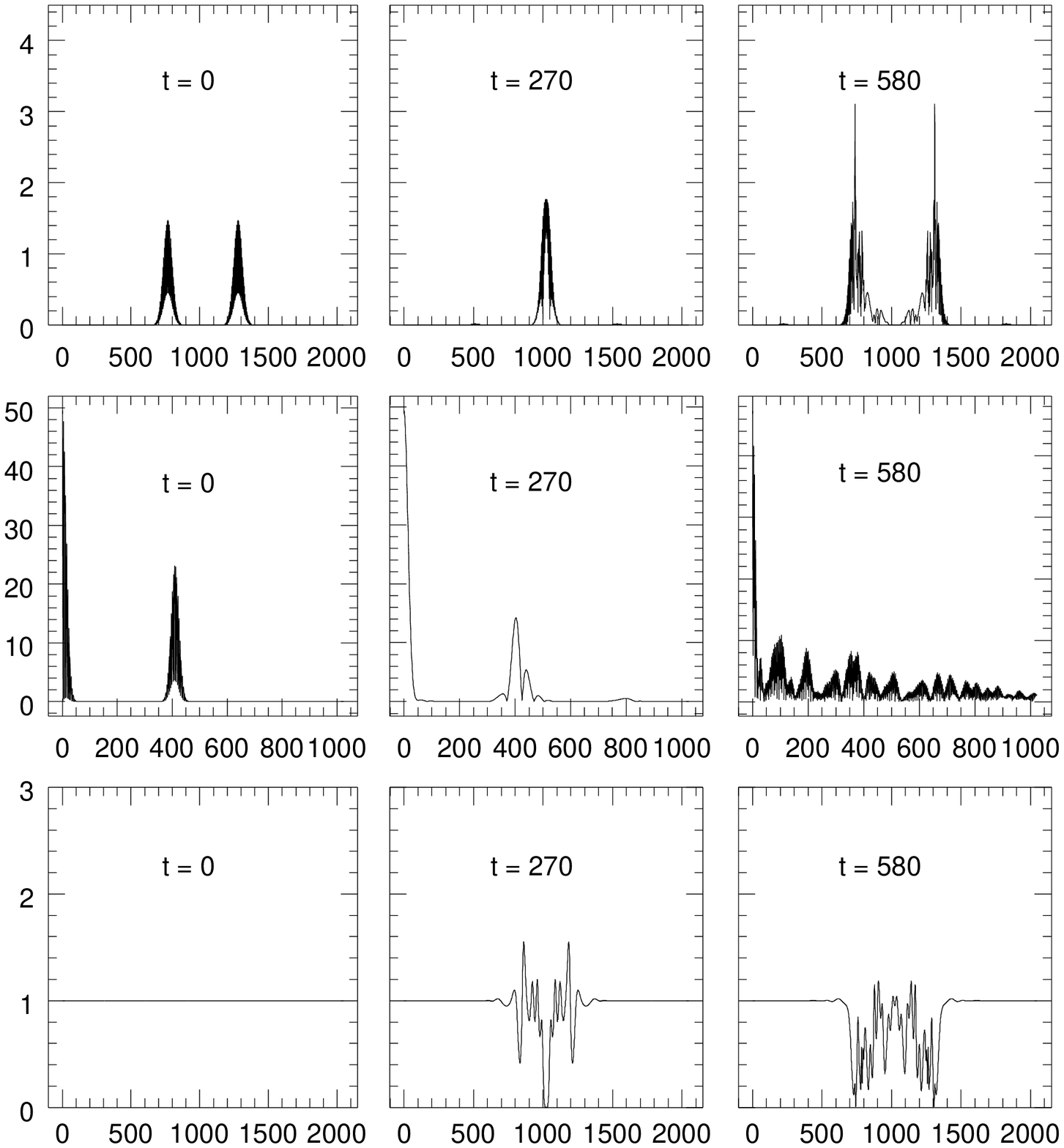}}
\centerline{Fig.2}
\else\fi

\ifx\pansw\pictures
\def\epsfsize#1#2{0.95#1}
\centerline{\epsfbox{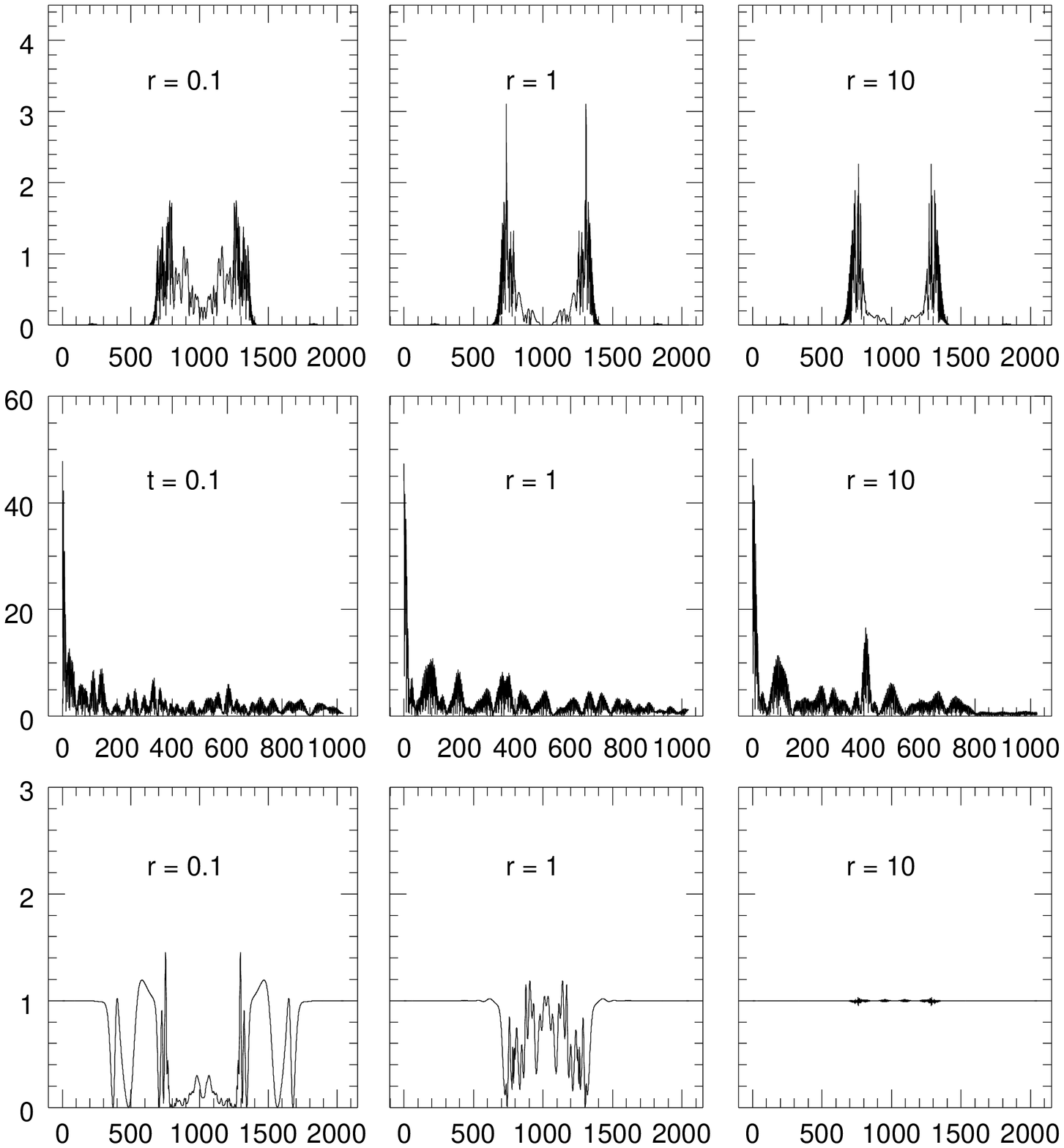}}
\centerline{Fig.3}
\else\fi

\ifx\pansw\pictures
\def\epsfsize#1#2{0.95#1}
\centerline{\epsfbox{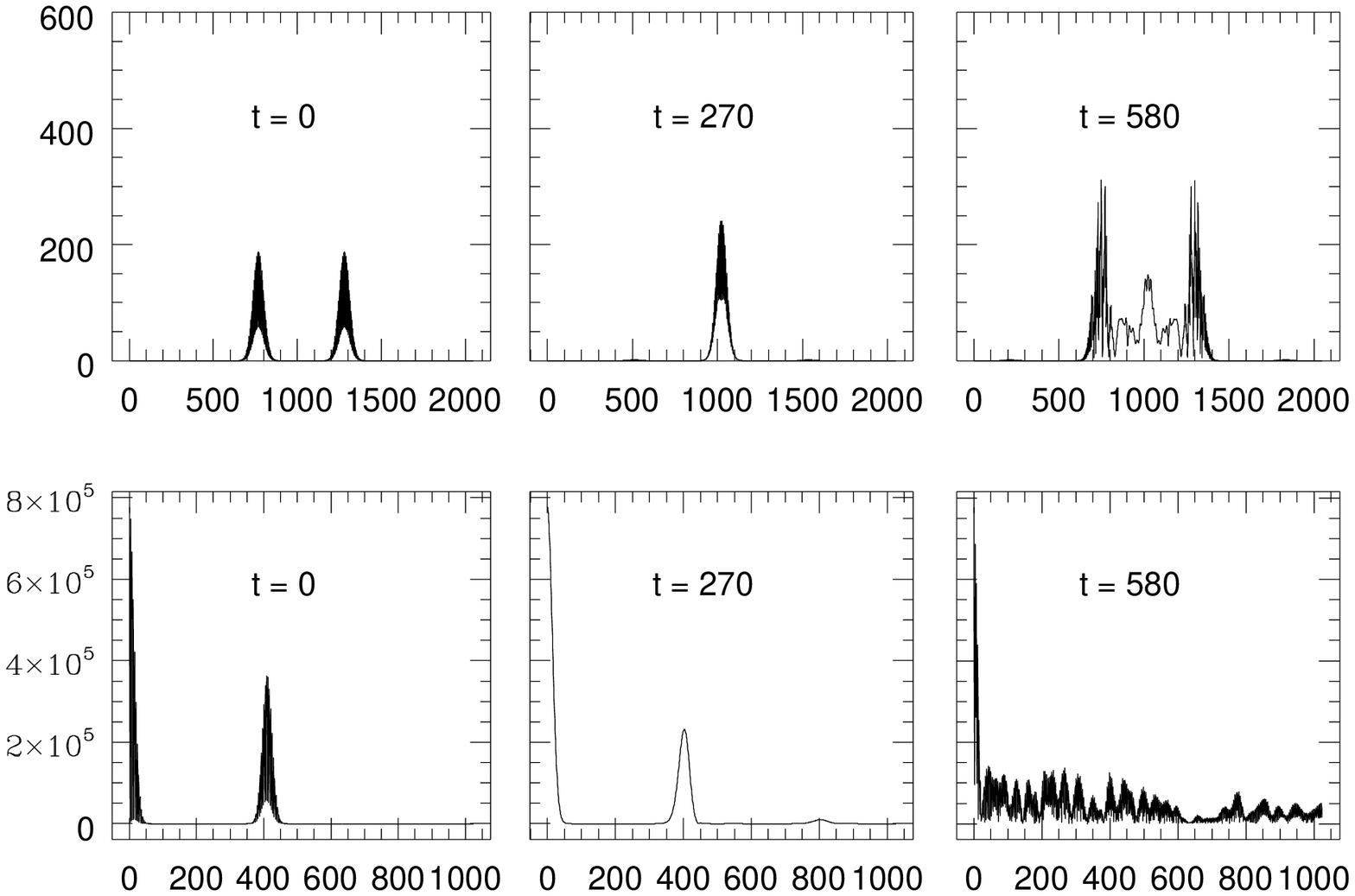}}
\centerline{Fig.4}
\else\fi

\ifx\pansw\pictures
\def\epsfsize#1#2{0.95#1}
\centerline{\epsfbox{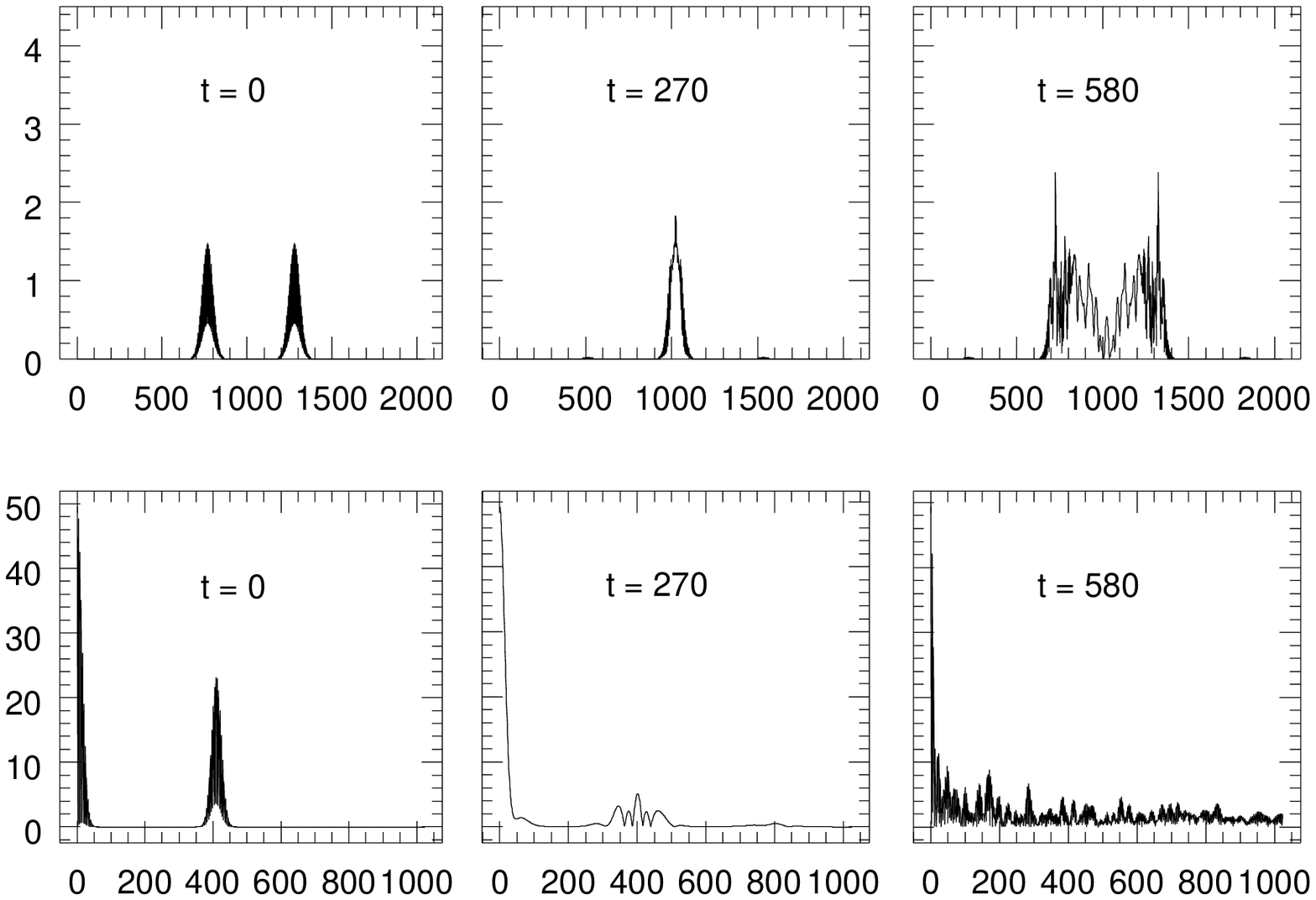}}
\centerline{Fig.5}
\else\fi

\ifx\pansw\pictures
\def\epsfsize#1#2{0.95#1}
\centerline{\epsfbox{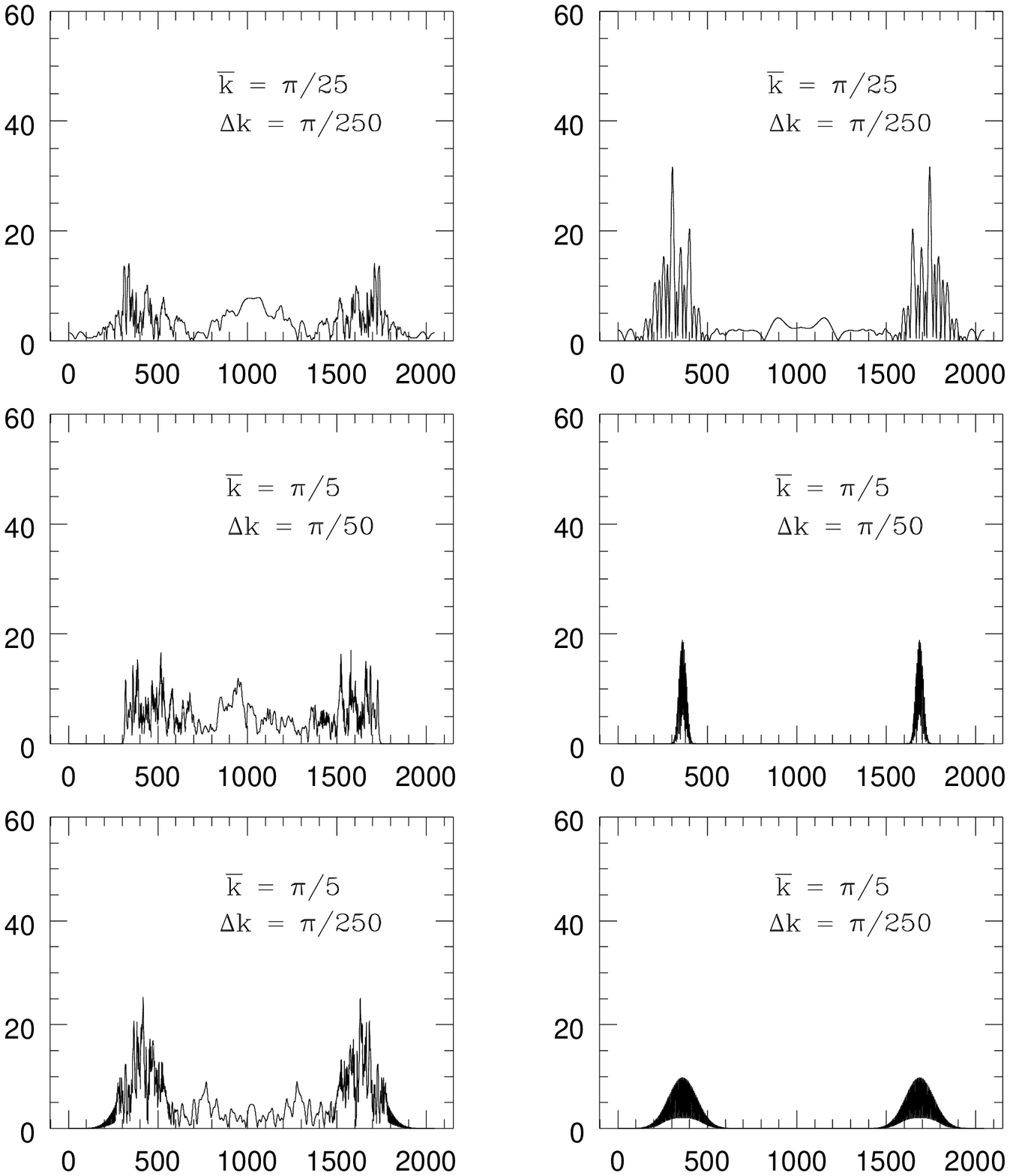}}
\centerline{Fig.6}
\else\fi

\ifx\pansw\pictures
\def\epsfsize#1#2{0.95#1}
\centerline{\epsfbox{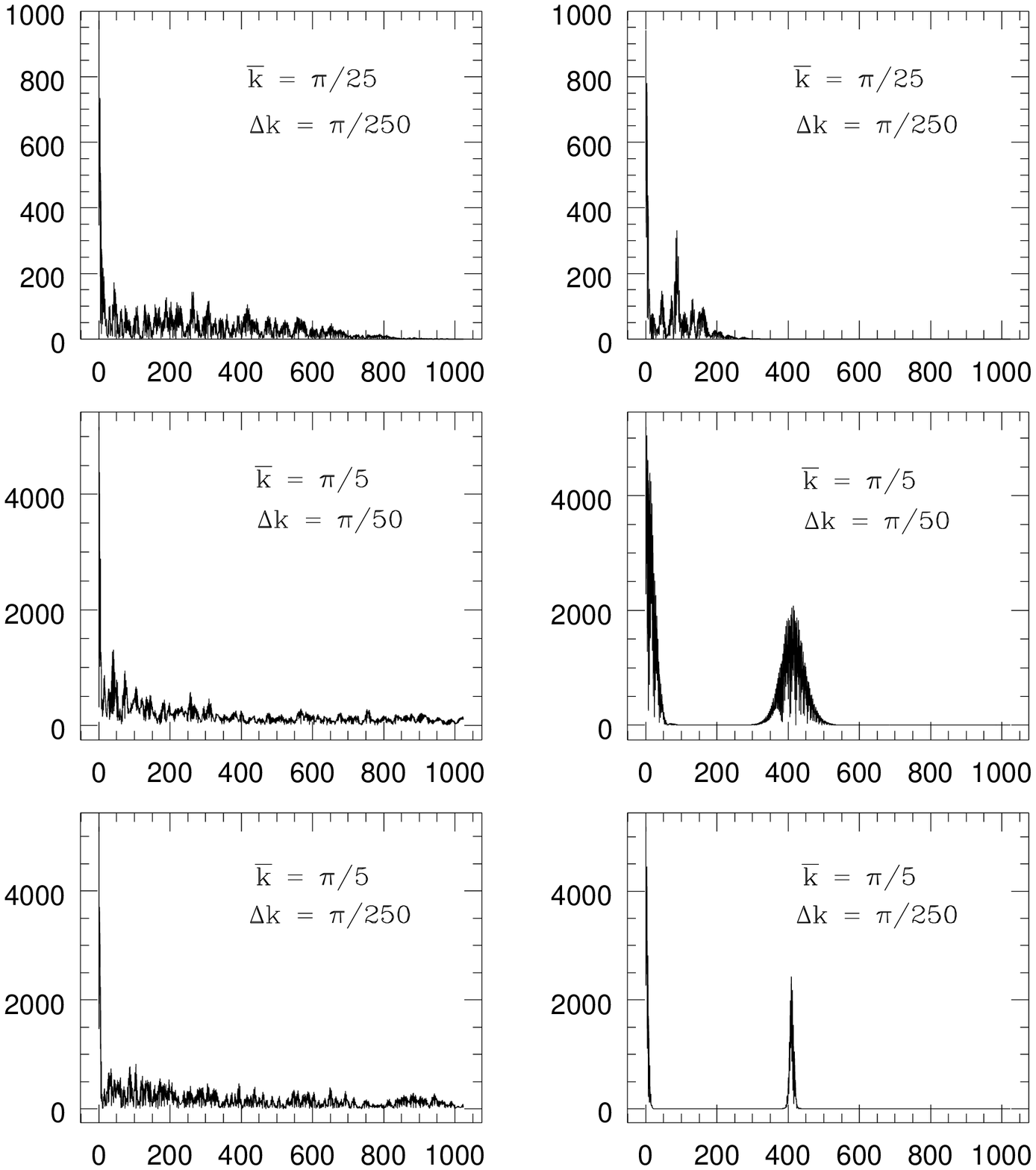}}
\centerline{Fig.7}
\else\fi

\end{document}